# Science Autonomy using Machine Learning for Astrobiology

*A white paper for the 2025 NASA Decadal Astrobiology Research and Exploration Strategy (DARES)*

**Lead Author:** Victoria Da Poian, Microtel LLC / Tyto Athene LLC at NASA Goddard Space Flight Center (GSFC), victoria.dapoian@nasa.gov

**Co-authors:** Bethany Theiling[1], Eric Lyness[1,2], David Burtt[1,3], Abigail R. Azari[4,5], Joey Pasterski[1,3], Luoth Chou[1], Melissa Trainer[1], Ryan Danell[1,6], Desmond Kaplan[1,6], Xiang Li[1], Lily Clough[1,7], Brett McKinney[7], Lukas Mandrake[8], Bill Diamond[9], Caroline Freissinet[12]

**Endorsers:** Scott D. Guzewich[1], Jamie Elsila[1], Jennifer L. Eigenbrode[1], Stephanie A. Getty[1], Ashley M. Hanna[10], Joanna Clark[11], Amy McAdam[1], Marie Farrell[13], Alessandro Pinto[8], Francesco Civilini[1], Umaa Rebbapragada[8], Jack Lightholder[8], Kendra K. Farnsworth[1,14], Conor A. Nixon[1]

**Affiliations**: 1. NASA Goddard Space Flight Center, 2. Microtel LLC / Tyto Athene LLC, 3. Oak Ridge Associated Universities, 4. University of Alberta, Edmonton, Alberta, Canada, 5. Alberta Machine Intelligence Institute, Edmonton, Alberta, Canada, 6. Danell Consulting Inc., 7. University of Tulsa, 8. Jet Propulsion Laboratory, 9. SETI Institute, San Jose, 10.University of Maryland, College Park, MD, 11. Texas State University - Amentum JETSII Contract, Houston TX, 77062, 12. Centre National de la Recherche Scientifique, France, 13. The University of Manchester, UK, 14. University of Maryland, Baltimore County.

1) **<u>Motivation for Science Autonomy</u>**

In the last few decades, artificial intelligence (AI) including machine learning (ML) have become essential for data analysis in space missions [1]. AI and ML enable rapid processing of large datasets, and offer advanced feature extraction and pattern recognition capabilities that deliver meaningful insights, enhancing human analysts' ability to identify correlations within complex, multi-variable datasets. This is especially needed for astrobiology, where models must distinguish complex biotic patterns from intricate abiotic backgrounds. As data volume outpaces the capacity for timely data analysis, AI and ML become essential for data processing. They could also prove invaluable for the complex data analysis that will accompany flight instruments' advancements. ML has been widely applied in image processing of large datasets in astrophysics and Earth observation (*e.g.*, crater identification [2-4], sample targeting [5]). Similar techniques that share methodology but are improved for onboard computational restrictions could be leveraged for astrobiology missions to identify key features [6]. This paper, primarily addressing the RFI's Topic 2 "Emerging Themes and Technologies", focuses on using onboard intelligence ('science autonomy') for mass spectrometry (MS) data analysis, a powerful chemical analysis technique with high life-detection potential [7, 8]. For more details on MS for astrobiology, see Pasterski *et al.* DARES submission.

As space missions venture to more distant planetary bodies, they face critical challenges such as fundamental communication limits (light travel time), mission design challenges (low bandwidth) and limited power/storage resources, further strained by the increasing data volume from advanced instruments. Missions traveling far from Earth (*e.g.,* Dragonfly, Europa Clipper) must operate under strict data transmission constraints, limiting data availability for science analysis. Continued investment in data return facilities (*e.g.,* Deep Space Network upgrades) and other technologies to enhance data return are required [1]. While infrastructure investments help to mitigate communication bottlenecks and maximize science return, AI and ML enable capabilities like onboard autonomy and ML-driven analysis beyond traditional infrastructures.

Our long-term vision for space missions involves *in situ* analysis, where spacecraft analyze data in real-time, make autonomous decisions, and prioritize scientific goals without relying solely on Earth-based instructions. Communication of explainable decisions by autonomous agents will remain crucial for accountability and feedback. While AI and ML tools have helped to mitigate deep-space missions communication latency, advancements in capability and the growing complexity of science instruments require commensurate improvements to our current ML techniques. These tools can enhance mission efficiency by optimizing data transmission, enabling opportunistic science (*e.g.,* Enceladus plumes [6]), detecting anomalies, and optimizing resources like energy allocation and scheduling. Section 2 **showcases ML functionality for backward-facing applications** on already-collected data, and Section 3 explores **forward-facing applications to enhance and enable future space missions**.

2) **Backward-facing applications**
   a) **Use case: Frontier Development Lab (FDL)**

   ML applications in planetary science have grown in recent years, with 3773 unique publications from Scopus in 2024 alone [1,9]. As part of efforts in ML for planetary science, researchers revisited datasets from NASA's Curiosity rover, which arrived in Gale Crater on Mars in 2012 [10]. Among the advanced scientific payload, the Sample Analysis at Mars (SAM) instrument suite and CheMin instrument play a crucial role in studying Martian geology and Mars' potential for supporting life [11]. Through a Frontier Development Lab (FDL) challenge over summer 2024, ML techniques were applied to previously-collected SAM dataset analysis, demonstrating that these methods could support scientists' decision-making during time-limited operations. These efforts centered on transfer learning, a ML technique where a pre-trained model is adapted to a related task [12]. Specifically, Da Poian *et al.* [12] investigated transfer learning between commercial and flight-like instruments (more data) with actual space missions instruments (less data) and showed that transferability between these datasets exists and could be leveraged to train ML algorithms for planetary science. Preparing and curating SAM and CheMin data revealed limitations in metadata collection and archiving, informing improvements for instruments in development, such as the ExoMars Mars Organic Molecule Analyzer suite and the Dragonfly Mass Spectrometer analytical lab on Dragonfly to explore Titan.

   b) **Strategies for Enabling ML**

   This FDL 2024 challenge revealed the difficulties in applying ML to NASA's Planetary Data System (PDS) archives. Much of the PDS-archived data is not in a ML-readable format, requiring additional effort to create curated, consistent, high-quality datasets for ML applications. While "ML-ready" is subjective, efforts by Laura *et al.* [13] and a Planetary Data Ecosystem review board [14] propose solutions for the planetary science community. ML applications face hurdles in data quality, formatting, and volume, as well as the difficulty of navigating the current PDS data systems (despite recent updates) without expert support. Mission instrument teams develop unique processes and software to optimize hardware performance, conserve limited lifetime resources, and mitigate hardware failure risks. Since these processes are developed during mission operations, key metadata is often missing from archived data, which is typically designed well before the operations phase.

   Incorporating a subsection of ML-readiness in the Data Management Plan section of ROSES proposals and mission announcements of opportunity could ensure that datasets are properly formatted, contain the necessary metadata, and are archived for future applications. These standards would need to be implemented *via* a top-down approach, where NASA headquarters (HQ) would develop standards with input from the planetary/ML/proposing community, and upon implementation, provide resources and guidance to proposers to accommodate various levels of ML familiarity and "regularly assess the Findability, Accessibility, Interoperability, and Reusability of data" [14].

Moreover, despite NASA's desire to drive innovation and research in AI and ML for space missions, the lack of dedicated and sustained solicitations and funding (*e.g.,* ROSES) hinders progress. For instance, the FDL 2024 challenge relied on **passionate volunteer** scientists, who contributed approximately ~48 hours each over eight months, equating to ~$17,000 (for 6 scientists, assuming a ~$60 hourly rate, without overhead costs). Direct solicitations would maximize the science return of existing and upcoming missions in a standardized and cost-effective manner rather than depending upon case-by-case volunteer work. The absence of allocated time and resources for subject matter experts slows the progress of making PDS data ML-ready, limiting the innovation and implementation necessary for outer solar system missions prioritized by NASA HQ.

3) <u>**Forward-facing applications**</u>
   a) <u>**Applications for future missions**</u>

AI and ML are transformative approaches for future missions, enabling expanded autonomous operations, real-time decision-making, choosing the most relevant data to send back to Earth in a low bandwidth case, and thus enhancing astrobiology investigations. Here we highlight key autonomy applications that will enhance and enable future missions in environments previously beyond our operational reach.

- **Onboard data analysis using trained algorithms:** Autonomous techniques for chemical biosignature identification (*e.g.*, pattern recognition, predictions for hypothesized signatures, anomaly detection) will need to operate robustly and efficiently under resource-limited conditions with a strong verification and validation pipeline to ensure reliability.

- **Autonomous decision-making** will be critical to optimize science on distant planetary bodies, enabling spacecraft to select high-priority samples in real-time, dynamically adapt mission objectives, and maximize science return [15]. Missions like Perseverance and Curiosity on Mars have demonstrated the potential of autonomous target selection and observation strategies (*e.g.*, AEGIS [16]), while future missions such as Dragonfly on Titan and those exploring ocean worlds will require even greater autonomy. Instrument-based autonomy also facilitates onboard collaboration between payloads, enhancing confidence in life detection through multiple complementary measurements. Moreover, these real-time analysis systems are essential for prioritizing onboard actions, like deciding whether to store data, transmit it to Earth, or collect additional data. These systems enable real-time tuning of instruments based on newly collected data, adapt science workflows in response to unexpected discoveries, and support opportunistic science, with various degrees of involvement from ground teams [17].

- **Collaborative science** involves integrating datasets from multiple instruments to provide a "full-picture" of the sample or environment. This requires coordination strategies from multiple instruments and potential for collaboration between different missions (*e.g.*, such as rovers and orbiters on the same target). While this collaboration is done with ground-in-the-loop (*e.g.,* SAM - CheMin on Curiosity), such advancements promise to enhance mission efficiency and scientific output (*e.g.,* [15,18]).

      **b) <u>Challenges</u>**

While AI and ML are rapidly evolving fields, space exploration missions rely on high-heritage technologies and processes. The challenge of integrating novel techniques into space missions appears in three areas - hardware, software, and process. These challenges are particularly relevant to astrobiology as these applications will need customized AI research tools and can't only rely on off-the-shelf solutions.

From a hardware perspective, the limited onboard computing power restricts onboard algorithms (*i.e.*, low computational intensive ML models). For instance, Curiosity uses a BAE RAD 750 processor from the 1990s, and ExoMars will be using the LEON 2 processor, both of which possess flight heritage. While more powerful computing boards exist, qualifying them for space missions is arduous (power/weight limitations, vibration and radiation exposure, and Technology Readiness Level advancement). For now, these computing limitations necessitate the use of low-computational-intensive ML models capable of functioning within the mission's constraints (*e.g.*, energy, communication, onboard computing limits). On the software side, spacecrafts' software **must** be reliable and robust, requiring fault-tolerant systems that can adapt to anomalies and protect the instruments and the mission integrity to ensure mission success in the face of uncertainty and extreme conditions.

Moreover, the application of AI and ML in space exploration also raises critical ethical and policy considerations. As we do not always know what we are looking for (*e.g.,* life as we don't know it), experts must carefully define questions and develop algorithms to address the unknown(s) that we are trying to unveil. Communication of explainable decisions by autonomous agents remains crucial for accountability and feedback. Mission stakeholders, more precisely scientists and ML engineers, bear the responsibility of ensuring the proper tools design to avoid misinterpretation, even more for life detection questions. In addition, trust in AI-driven algorithms, especially for life detection missions, remains a major obstacle. A key concern is that ML models are often deployed as black-boxes, making it challenging for scientists to rely on them fully, especially for such groundbreaking discoveries. To address this, investment in transparent and explainable model approaches is critical, to ensure AI-driven tools provide interpretable outputs and include false positive diagnostics [19]. While many aspects of future missions will be autonomous, the final biosignature's interpretation will remain a human responsibility. Building trust requires integrating AI tools very early in the mission conception and development process - a shift from the traditional paradigm that will necessitate cultural and procedural changes in space mission development.

### 4) Recommendations

Advancing the integration of autonomy through AI and ML into space missions is a complex challenge, and we believe that by focusing on key areas, we can make significant progress and offer practical recommendations for tackling these obstacles.

First, we recommend NASA and industry partners continue to push towards the **development of hardware capable of allowing real-time AI computations onboard spacecraft** (High Performance Spaceflight Computing) as it is a critical step toward advancing autonomy onboard space missions. Also, **advancing data transmission tools** such as laser communication systems (*e.g.*, ILLUMA-T instrument), and **continuing investments in the Deep Space Network** will further enhance the efficiency of data transmission of future missions.

Second, from a data and algorithms perspective, having a consistent and robust **intelligent data processing pipeline** that includes data collection, data labeling, data analysis, and data management systems is compulsory for applications of ML algorithms. Our team members at NASA GSFC have built expertise in this process for MS data and are writing a framework strategy to share with the community that could be leveraged for the "data science plan" that we recommend for implementation in proposals' "data management plan". Moreover, we need to continue to develop **ML models for dedicated scientific purposes** (*e.g.,* anomaly detection, sensor failures predictions, science data analysis) with models trained on Earth and fine-tuned for specific mission targets. As mentioned above, the disparity between the rapid pace of AI innovation and NASA's slower development processes emphasizes the need for flexible frameworks and ML benchmarks, particularly in astrobiology-related models. These benchmarks will ideally leverage specific datasets and test various models to investigate NASA's main science questions (including astrobiology, planetary science, Earth science). Finally, we recommend **NASA organize dedicated funding programs for the investigation of AI and ML tools for science data analysis**. It would ultimately save NASA time, money, and resources while improving missions' efficiency. In these solicitations, we recommend requirements for standardized data formatting and metadata collection processes to ensure accessibility and AI-ready dataset archiving.

Finally, in order for stakeholders to trust the implementation of autonomy, we recommend developing consistent, step-by-step testing of AI and ML methods through rigorous applications in the lab and analog field settings. Our main recommendation is to have involvement of scientists, engineers, mission management, software developers, and data scientists from the beginning of mission concept development.